\documentclass[aps,prl,twocolumn,10pt,showpacs,superscriptaddress,amsmath,amssymb,nobibnotes]{revtex4-1}

\usepackage{graphicx}
\usepackage{amsfonts}    
\usepackage{graphicx}   
\usepackage{verbatim}   
\usepackage{color}      
\usepackage{subfigure}  
\usepackage{hyperref}   
\usepackage{natbib}
\usepackage{booktabs}
\usepackage{threeparttable}
\usepackage{dcolumn}
\usepackage{multirow}

\begin{document}
\title{Gravitational Waves Detection via Weak Measurements Amplification}
\author{Meng-Jun Hu}\email{mengjun@mail.ustc.edu.cn}
\author{Yong-Sheng Zhang}\email{yshzhang@ustc.edu.cn}
\affiliation{Laboratory of Quantum Information, University of Science and Technology of China, Hefei 230026, China}
\affiliation{Synergetic Innovation Center of Quantum Information and Quantum Physics, University of Science and Technology of China, Hefei 230026, China}

\date{\today}

\begin{abstract}
A universal amplification scheme of ultra-small phase based on weak measurements is given and a weak measurements amplification based laser interferometer gravitational-wave observatory (WMA-LIGO) is suggested. The WMA-LIGO has potential to amplify the ultra-small phase signal to at least $10^{3}$ order of magnitude such that the sensitivity and bandwidth of gravitational-wave detector can be further improved. Our results not only shed a new light on the quantum measurement but also open a new way to the gravitational-wave detection.  
\end{abstract}

\maketitle
{\it Introduction.}
The most exciting and important scientific event in the beginning of year 2016 undoubtedly should be the direct observation of gravitational waves by advanced Laser Interferometer Gravitational-Wave Observatory (LIGO) \cite{1, second}. The detection of ultra-small gravitational waves by high sensitivity LIGO \cite{2} implies the new era of gravitational-wave astrophysics. The sensitivity of gravitational-wave detectors can be further improved with the latest state-of-art technologies \cite{improve1,improve2} and the possible global gravitational-wave network joined by Advanced Virgo \cite{3}, the KAGRA interferometer \cite{4} and the planned third LIGO in India \cite{5} will tremendously enhances parameter estimation and sky location \cite{location}. The LIGO detector is essentially a modified Michelson interferometer, in which phase difference between its two orthogonal arms caused by gravitational-wave strain is measured \cite{Weiss}. Except for using significant developed technology and new facilities to dramatically improve the sensitivity of gravitational-wave detectors, it is interesting and important to ask whether or not there exists possible types of gravitational-wave detectors with high sensitivity other than the type of  Michelso interferometer, which operate based on different theory of measurement. The recent progress in the theory of quantum measurement e.g., weak measurements may offer positive answer.

Weak measurement, which was proposed by Aharonov, Albert and Vaidman (AAV), focus on disturbing system as small as possible so that the state of a system would not collapse after measurement and a projective measurement is subsequently performed for realizing post-selection of the state of the system \cite{AAV}. In AAV theory, the mystery weak value $\langle\hat{A}\rangle_{w} \equiv \langle\psi_{i}|\hat{A}|\psi_{f}\rangle/ \langle\psi_{i}|\psi_{f}\rangle$ of observable $\hat{A}$, with $|\psi_{i}\rangle$ and $|\psi_{f}\rangle$ are pre-selected and post-selected state respectively, which can be complex and far beyond spectra of $\hat{A}$, emerges when von Neumann-type interaction is considered. Despite of its ambiguous physical meaning \cite{Fe, Com, Dre}, weak value has been shown powerful in solving paradox \cite{para}, reconstructing quantum state \cite{state1, state2, state3}, and especially amplifying ultra-small signal \cite{am1, am2, am3}. When $\langle\psi_{i}|\psi_{f}\rangle\rightarrow 0$, the weak value $\langle\hat{A}\rangle_{w}$ becomes extremely large but with very low success probability $P\approx|\langle\psi_{i}|\psi_{f}\rangle|^{2}$. The obtained expectated value of a proper chosen pointer observable is proportional to $\theta\langle\hat{A}\rangle_{w}$, where $\theta$ is the signal need to be measured and manifests itself as the coupling in the interaction Hamiltonian \cite{Josz,Dres}. The weak value amplification (WVA), which is at the cost of low success probability of post-selection, has been focused of argument on whether WVA measurement outperforms conventional measurement in parameter estimation recently \cite{out1,out2,out3}. However, the fact that WVA measurement has potential to magnify ultra-small signal to the measurable level undoubtedly makes it powerful and effective. For instance, the thickness of one paper is obviously much smaller than the minimum attainable value of a regular ruler. With one hundred of papers together, however, we can not only estimate the thickness of one paper but also obtain precision one hundred higher than ruler itself. Besides, the WVA measurement has been shown that it can perform better than conventional measurement in the case of existence of some types of technical noises \cite{tech1,tech2,tech3}. 

The possibility of realizing ultra-small phase measurement via WVA is intriguing and will significantly promote the development of precision measurement. Although special phase measurement has been investigated \cite{am3,phase2,phase3}, a general phase measurement protocol based on interferometric WVA still seems infeasible in the practical operation \cite{phase1}. The reason is that the pointer i.e., frequency of laser must be distributed so wide that the condition of weak measurements is satisfied. The laser pulse, in this case, must be very short with limited coherence length. It will be technically difficulty to build an interferometer with this kind of laser source. Besides, it is impossible to guarantee the perfect Gaussian distribution of frequency of laser. 

In this letter, we propose a universal amplification measurement scheme of ultra-small phase based on weak measurements. Contrary to WVA, the amplification factor of our scheme is not determined by the weak value, which has no explicit definition in our case. A weak measurements amplification based laser interferometer gravitational-wave detector is then suggested here and shown that it has potential to further enhance the sensitivity and broaden the bandwidth of gravitational-wave detectors.

{\it Ultra-small phase amplification via weak measurements.}  
We begin with a brief review of the basic idea of WVA and then introduce the weak measurement amplification of ultra-small phase. Consider a two-level system initially prepared in a state of superposition $|\psi_{i}\rangle=\alpha|0\rangle+\beta|1\rangle$ with $|\alpha|^{2}+|\beta|^{2}=1$. Unlike continuous pointer in the most discussions \cite{AAV}, here we choose discrete pointer e.g., polarization degree of freedom of photon for practical purpose. The initial state of composite system becomes $|\Psi_{i}\rangle_{SP}=|\psi_{i}\rangle\otimes|H\rangle$ in which $|H\rangle$ representing horizontal state of polarization is the initial state of pointer. The interaction Hamiltonian of von Neumann-type takes the form $\hat{H}=g\hat{A}\otimes\hat{\sigma}_{y}$ with $\hat{A}\equiv |0\rangle\langle 0|-|1\rangle\langle 1|$ is the observable of system and $\hat{\sigma}_{y}$ is the Pauli operator \cite{state1}. Different eigenstate of observable $\hat{A}$, after interaction, causes different rotation of the pointer. When the system is post-selected in the state $|\psi_{f}\rangle=\gamma|0\rangle+\eta|1\rangle$ with $|\gamma|^{2}+|\eta|^{2}=1$, the state of pointer becomes (unnormalized)
\begin{equation}
|\tilde{\varphi}\rangle_{P}=\langle\psi_{f}|e^{-i\theta\hat{A}\otimes\hat{\sigma}_{y}}|\psi_{i}\rangle\otimes|H\rangle,
\end{equation}
where natural unit is used i.e., $\hbar\equiv 1$ and $\theta=g\Delta t$ with $\Delta t$ is the time of interaction. If the interaction is weak enough i.e., $\theta\ll 1$, the state of the pointer, in the first order approximation, can be rewritten as
\begin{equation}
|\tilde{\varphi}\rangle_{P}=\langle\psi_{f}|\psi_{i}\rangle e^{-i\theta\langle\hat{A}\rangle_{w}\hat{\sigma}_{y}}|H\rangle,
\end{equation}
where $\langle\hat{A}\rangle_{w}=\langle\psi_{f}|\hat{A}|\psi_{i}\rangle/\langle\psi_{f}|\psi_{i}\rangle$ is the weak value of $\hat{A}$. The post-selection of the system in the case of weak coupling thus causes $\theta\langle\hat{A}\rangle_{w}$ rotation of pointer. The small signal $\theta$ can be extracted by performing a proper observable measurement on the pointer, specifically we have
\begin{equation}
\begin{split}
\langle\hat{\sigma}_{+}\rangle_{P}&=2\theta\mathrm{Re}\langle\hat{A}\rangle_{w} \\
\langle\hat{\sigma}_{R}\rangle_{P}&=2\theta\mathrm{Im}\langle\hat{A}\rangle_{w},
\end{split}
\end{equation}
where $\hat{\sigma}_{+}\equiv |+\rangle\langle +|-|-\rangle\langle -|$, $\hat{\sigma}_{R}\equiv |R\rangle\langle R|-|L\rangle\langle L|$ and $|\pm\rangle=(|H\rangle \pm|V\rangle)/\sqrt{2}$, $|R\rangle=(|H\rangle+i|V\rangle)/\sqrt{2}$, $|L\rangle=(|H\rangle-i|V\rangle)/\sqrt{2}$ with $|V\rangle$ represents state of vertical polarization. The amplification factor, which determined only by weak value $\langle\hat{A}\rangle_{w}$, can be very large when the post-selected state $|\psi_{f}\rangle$ is almost orthogonal to the initial state $|\psi_{i}\rangle$ of system. 

As we have stated, it is impossible to realize general ultra-small phase amplification by using the above WVA method. To obtain amplification of ultra-small phase $\theta$, we choose $|+\rangle$ as the initial state of pointer and consider the unitary operator of evolution
\begin{equation}
 \hat{U}=|0\rangle\langle 0|\otimes \hat{I}+|1\rangle\langle 1|\otimes(|H\rangle\langle H|+e^{i\theta}|V\rangle\langle V|),
\end{equation} 
which evolves the composite system to the state of
\begin{equation}
|\Psi_{f}\rangle_{SP}=\alpha|0\rangle\otimes|+\rangle+\beta|1\rangle\otimes(|H\rangle+e^{i\theta}|V\rangle)/\sqrt{2}.
\end{equation}
It is a control-rotation in which the initial state of the pointer is rotated $\theta$ along the equator of the Bloch sphere when the system is in the state $|1\rangle$ and nothing happens when the system is in the state $|0\rangle$. After post-selection, the state of the pointer becomes (unnormalized)
\begin{equation}
 |\tilde{\varphi}\rangle_{P}=_{S}\langle\psi_{f}|\Psi_{f}\rangle_{SP}=(\alpha\gamma+\beta\eta)|H\rangle+(\alpha\gamma+\beta\eta e^{i\theta})|V\rangle
\end{equation} 
with $\alpha,\beta,\gamma,\eta$ are taken real numbers here. We can recast $\alpha\gamma+\beta\eta e^{i\theta}$ as $\sqrt{\alpha^{2}\gamma^{2}+\beta^{2}\eta^{2}+2\alpha\gamma\beta\eta\mathrm{cos}\theta}e^{i\phi}$ with 
\begin{equation}
\mathrm{tan}\phi=\dfrac{\beta\eta\mathrm{sin}\theta}{\beta\eta\mathrm{cos}\theta+\alpha\gamma}.
\end{equation}
Since phase signal $\theta\ll 1$, $\sqrt{\alpha^{2}\gamma^{2}+\beta^{2}\eta^{2}+2\alpha\gamma\beta\eta\mathrm{cos}\theta}=\alpha\gamma+\beta\eta$ in the first order approximation and the state of the pointer thus becomes
\begin{equation}
 |\varphi\rangle_{P}=\dfrac{1}{\sqrt{2}}(|H\rangle+e^{i\phi}|V\rangle).
\end{equation} 
The operation of post-selection after weak interaction results in $\phi$ rotation of pointer state along equator of Bloch sphere. The amplification can be realized when the post-selected state $|\psi_{f}\rangle$ is properly chosen such that condition $\phi=\mathrm{arctan}[\beta\eta\mathrm{sin}\theta/(\alpha\gamma+\beta\eta\mathrm{cos}\theta)]>\theta$ is satisfied.   For example, when $\alpha=\beta=1/\sqrt{2}$ and $\gamma=\mathrm{cos}\chi$, $\eta=\mathrm{sin}\chi$, in the case of $\theta\ll 1$, Eq. (7) becomes
\begin{equation}
\mathrm{tan}\phi=\dfrac{\theta}{1+\mathrm{cot}\chi}.
\end{equation}
To obtain significant amplification, $1+\mathrm{cot}\chi\ll 1$ is required. Suppose $\chi=-(\pi/4+\delta)$ with $\delta\ll 1$, then $\mathrm{cot}\chi=-1+\delta$ in the first order approximation and we have $\mathrm{tan}\phi=\theta/\delta$. The factor of amplification in this case is 
\begin{equation}
h=\dfrac{\phi}{\theta}=\dfrac{\mathrm{arctan}(\theta/\delta)}{\theta}.
\end{equation}
The amplified phase signal $\phi$ shown in the final state $|\varphi\rangle_{P}$ can be extracted by performing proper measurement e.g., $\lbrace|R\rangle, |L\rangle\rbrace$ basis measurement on the pointer and the resulting differential probability is 
\begin{equation}
p=\mathrm{sin}\phi.
\end{equation}

Our amplification method, in analogy to micrometer that transforms a small displacement into a larger rotation of circle, transduces a ultra-small phase signal into a larger rotation of the pointer state along equator in the Bloch sphere.

\begin{figure*}[tbp]
\centering
\includegraphics[scale=0.5]{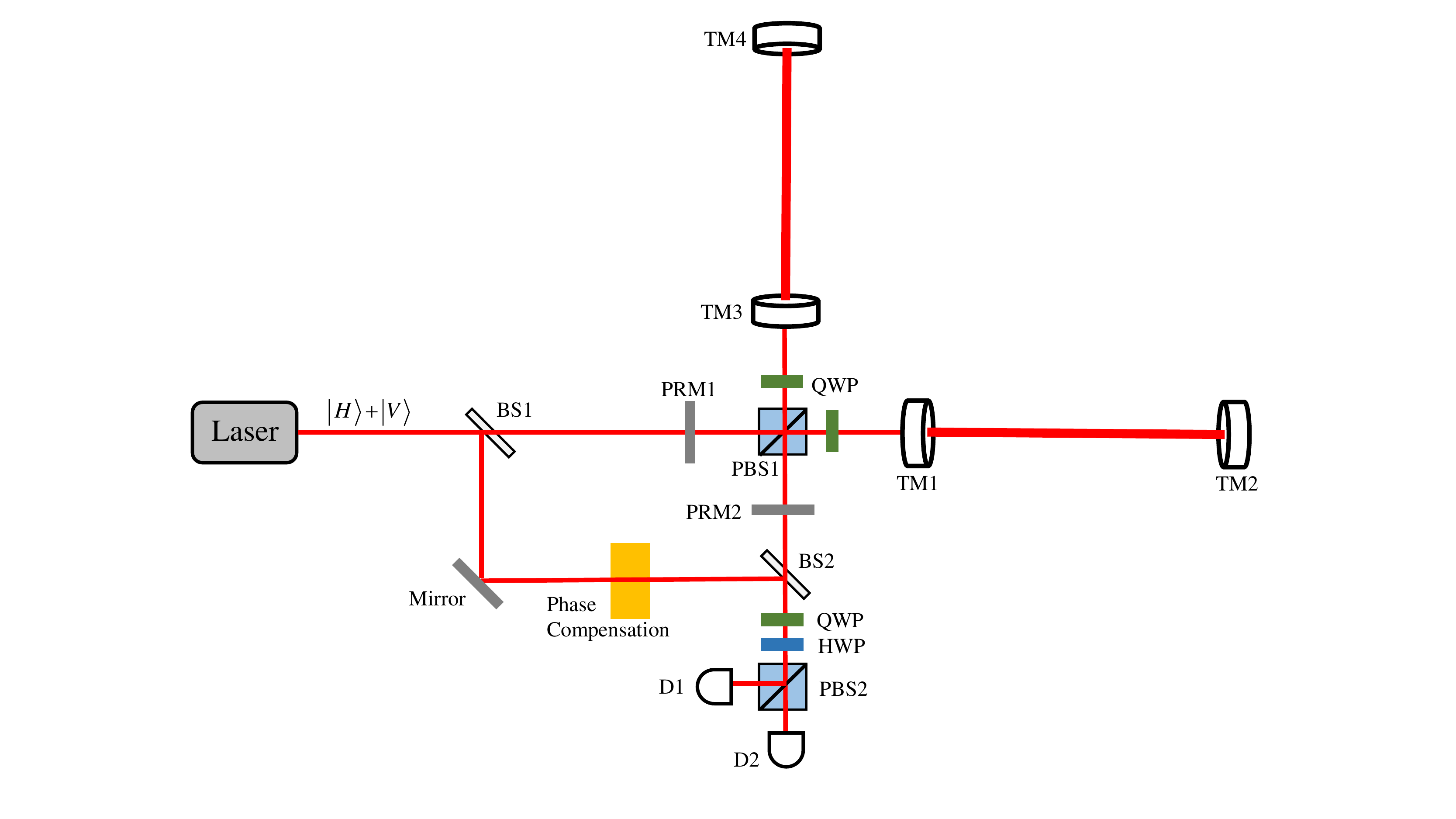}
\caption{Schematic diagram of Weak Measurements Amplification based Laser Interferometer Gravitational-wave Observatory (WMA-LIGO). BS: beam splitter, PBS: polarizing beam splitter, PRM: power recycle mirror, HWP: half wave plate, QWP: quarter wave plate, TM: test mass, D: detector. The coefficient of reflection and transmission of BS1 and BS2 are $r_{1}, t_{1}$ and $r_{2}, t_{2}$ respectively. Power recycle is realized by PRM1 and PRM2 with same reflectivity. QWPs in the two arms are fixed at $\pi/4$ and QWP, HWP, PBS2, D1, D2 constitute a polarization analyzer. }
\end{figure*}

{\it Weak measurements amplification based gravitational-wave detector.}
The Fig. 1 shows the schematic diagram of Weak Measurements Amplification based Laser Interferometer Gravitational-Wave Observatory (WMA-LIGO). The WMA-LIGO, which operates on a different theoretical basis, keeps the key feature of LIGO i.e., two orthogonal arms . The similar configuration makes the future possible upgrade on the LIGO is feasible.

The WMA-LIGO consists of five parts i.e., laser source, initial state preparation, signal collection, signal amplification and signal detection. The laser source produces stabilized photons in the linear polarization state $|+\rangle=(|H\rangle+|V\rangle)/\sqrt{2}$. The initial state preparation is fulfilled by the beam splitter (BS1). The state of photons, after passing through the BS1, becomes
\begin{equation}
 |\Psi_{i}\rangle_{SP}=|\psi_{i}\rangle_{S}\otimes|+\rangle_{P}=(r_{1}|down\rangle+t_{1}|up\rangle)\otimes|+\rangle_{P},
 \end{equation} 
where $r_{1}, t_{1}$ are coefficients of reflection and transmission of the BS1 satisfying $|r_{1}|^{2}+|t_{1}|^{2}=1$ and $|up\rangle,|down\rangle$ represent path state of up arm and down arm respectively. The photons, which fly along path of up arm, enters a modified polarizing Michelson interferometer (PMI) that used for signal collection. The modified PMI, which consists of a polarizing beam splitter (PBS1), two quarter wave plates (QWP) and four test masses, outputs the state of photons $(|H\rangle+e^{i\theta}|V\rangle)/\sqrt{2}$ when the input state is $(|H\rangle+|V\rangle)/\sqrt{2}$ with $\theta$ is the phase signal to be measured. The function of QWP, which is fixed at $\pi/4$, is to transform polarization state $|H\rangle$ and $|V\rangle$ into its orthogonal state $|V\rangle$ and $|H\rangle$ respectively when photons pass through it twice. Hence the transmitted photons with the state $|H\rangle$ is converted by QWP to the state $|V\rangle$ , which is reflected by the PBS1, and the reflected photons with the state $|V\rangle$ is converted to the state $|H\rangle$ by the other QWP and is thus transmitted by the PBS1 such that the photons will not come out from the input port. The interaction of weak measurements is fulfilled when photons comes out of PMI with state
\begin{equation}
|\Psi_{f}\rangle_{SP}=r_{1}|down\rangle\otimes|+\rangle+t_{1}|up\rangle\otimes(|H\rangle+e^{i\theta}|V\rangle)/\sqrt{2}.
\end{equation}
The process of signal amplification, which depends on post-selection, can be fulfilled by the BS2 with coefficients of reflection and transmission are $r_{2}$ and $t_{2}$, which satisfy $|r_{2}|^{2}+|t_{2}|^{2}=1$. The post-selection is completed when we focus only on the photons comes out from the down port of BS2 with the post-selected path state
\begin{equation}
|\psi_{f}\rangle_{S}=r_{2}|down\rangle+t_{2}|up\rangle.
\end{equation}
The polarization state of post-selected photons, in the first order approximation, thus becomes
\begin{equation}
|\varphi\rangle_{P}=\dfrac{1}{\sqrt{2}}(|H\rangle+e^{i\phi}|V\rangle)
\end{equation}
with $\phi$ determined by
\begin{equation}
\mathrm{tan}\phi=\dfrac{\mathrm{sin}\theta}{\mathrm{cos}\theta+r_{1}r_{2}/t_{1}t_{2}}.
\end{equation}
The amplified phase signal $\phi$ thus can be obtained by properly choosing $r_{1}, r_{2}, t_{1}, t_{2}$ such that $_{S}\langle\psi_{f}|\psi_{i}\rangle_{S}=r_{1}r_{2}+t_{1}t_{2}\rightarrow 0$. The HWP, PBS2 and two detectors complete the detection of amplified phase signal as a polarization analyser. According to Eq. (10), the factor of amplification of WMA-LIGO only determined by $\delta$ and $\delta$ depends on the precision of optical elements. With current technology, $\delta$ can reach at least $10^{-3}$ order of magnitude and thus the factor of amplification $h\approx 10^{3}$.
 Note that phase compensation is needed in the down arm in order to obtain perfect interference in the BS2.

{\it Quantum noise analysis.}
The noise power spectrum direct determines the sensitivity of gravitational-wave detector. In order to detect gravitational-wave signal, noises such as photon shot noise and thermal noise must be reduced to acceptable level. Fortunately, the current noises reduced technologies used in the LIGO can also be used in the WMA-LIGO.

The sensitivity of gravitational-wave detectors is ultimately limited by quantum noises i.e., radiation-pressure noise and shot noise, which arise essentially from the quantum fluctuation of the vacuum field in the dark port side of beamsplitter  \cite{caves1,caves2}. While the radiation-pressure noise is proportional to the square root of input light power, the shot noise is inversely proportional to it and together they determine the minimum detectable gravitational-wave signal i.e., standard quantum limit \cite{kimble, chen}. We now consider the quantum noise in the case of WMA-LIGO with the assumption that all the optical elements are ideal for the simplicity of discussion without lossing validity. The core of WMA-LIGO is a polarizing Michelson interferometer, which implies that the quantum noise analysis is analogue to that of conventional Michelson interferometer \cite{caves1,caves2} except that the polarization of light field should be taken into consideration.   
Suppose that the input light field is in the Fock state with $|+\rangle$ polarization denoted by $|N,+\rangle$, the calculations give the radiation-pressure noise and shot noise as
\begin{equation}
\begin{split}
h_{rn}&\propto t_{1}\sqrt{N} \\
h_{sn}&\propto \dfrac{\sqrt{2}}{t_{1}t_{2}}\dfrac{1}{\sqrt{N}},
\end{split}
\end{equation}
where the condition of phase signal amplification $t_{1}t_{2}+r_{1}r_{2}\rightarrow 0$ is applied. Since that radiation-pressure noise is dominant in the lower frequency detection, the process of signal amplification via weak measurements, with the sacrifice in the higher frequency area, could improve the sensitivity of detector in the lower frequency area. 

{\it Discussion and Conclusion.}
The technical details of WMA-LIGO need further analysis and prototypes are required to do the full assessment of its noise power spectrum in the next step. Although WMA-LIGO is proposed for the gravitational-wave detection, it can also be applied in all kinds of high-precision measurements due to its ability to amplify the ultra-small phase.

In conclusion, we have proposed a universal amplification measurement scheme of ultra-small phase based on weak measurements. Based on this measurement scheme, a weak measurement amplification based gravitational-wave detector i.e., WMA-LIGO is suggested and shown that it has potential to improve the sensitivity and broaden bandwidth further due to its ability to amplify the phase signal. Considering current technology, at least $10^{3}$ order of magnitude amplification can be reached. Combined with Sagnac topology \cite{sag1,sag2}, it also has potential to realize quantum nondemolition measurement to beat standard quantum limit \cite{sag3}. Our results may open a new way to the gravitational-wave detection.

This work was supported by the National Natural Science Foundation of China (No. 11674306 and No. 61590932), the Strategic Priority Research Program (B) of the Chinese Academy of Sciences (No. XDB01030200) and National key R$\&$D program (No. 2016YFA0301300 and No. 2016YFA0301700).

\end{document}